\documentclass[%
reprint,
showpacs,preprintnumbers,
nofootinbib,
amsmath,amssymb,
aps,
prd,
floatfix,
]{revtex4-1}
\pdfoutput=1
\usepackage{graphicx}
\usepackage{dcolumn}
\usepackage{bm}
\usepackage{subfigure}
\usepackage{url}
\usepackage[hyperindex]{hyperref}
\newcommand{\nua}[1]{\ensuremath{\rlap{\kern-2.5pt\ensuremath{\overset{\scriptscriptstyle(-)}{\phantom{\nu}}}}{\ensuremath{{\nu}_{#1}}}}}
\begin{document}

\title{Appearance-Disappearance Relation in 3+$N_{s}$ Short-Baseline Neutrino Oscillations}

\author{C. Giunti$^{1}$}
\author{E. M. Zavanin$^{1,2,3}$}
\affiliation{
$^1$
INFN, Sezione di Torino, Via P. Giuria 1, I--10125 Torino, Italy
\\
$^2$
Department of Physics, University of Torino, Via P. Giuria 1, I--10125 Torino, Italy
\\
$^3$
Instituto de F\'\i sica Gleb Wataghin, Universidade Estadual de Campinas - UNICAMP,
\\
Rua S\'ergio Buarque de Holanda, 777, 13083-859 Campinas SP Brazil
}

\date{14 September 2015}

\begin{abstract}
We derive the relation between the
amplitudes of short-baseline appearance and disappearance oscillations
in 3+$N_{s}$ neutrino mixing schemes
which is the origin of the appearance-disappearance tension
that is found from the analysis of the existing data
in any 3+$N_{s}$ neutrino mixing scheme.
We illustrate the power of the relation
to reveal the appearance-disappearance tension
in the cases of
3+1
and
3+2
mixing
using the results of global fits of short-baseline neutrino oscillation data.
\end{abstract}


\maketitle

The standard three-neutrino mixing scheme,
which explains the oscillations that have been observed in
solar, reactor and accelerator experiments
(see \cite{Bellini:2013wra,Wang:2015rma}),
may be an incomplete description of neutrino mixing,
as suggested by the indications in favor of short-baseline oscillations
generated by a squared-mass difference
$\Delta{m}^2_{\text{SBL}} \sim 1 \, \text{eV}^2$
which is much larger than the standard
``solar''
($\Delta{m}^2_{\text{SOL}} \approx 7.5 \times 10^{-5} \, \text{eV}^2$)
and
``atmospheric''
($\Delta{m}^2_{\text{ATM}} \approx 2.4 \times 10^{-3} \, \text{eV}^2$)
squared-mass differences characteristic of three-neutrino mixing.

The indications in favor of short-baseline oscillations are:
the reactor antineutrino anomaly
\cite{Mention:2011rk},
which is a deficit of the rate of $\bar\nu_{e}$ observed in several
short-baseline reactor neutrino experiments
in comparison with that expected from the latest calculation of
the reactor neutrino fluxes
\cite{Mueller:2011nm,Huber:2011wv};
the Gallium neutrino anomaly
\cite{Abdurashitov:2005tb,Laveder:2007zz,Giunti:2006bj,Giunti:2010zu,Giunti:2012tn},
consisting in a short-baseline disappearance of $\nu_{e}$
measured in the
Gallium radioactive source experiments
GALLEX
\cite{Kaether:2010ag}
and
SAGE
\cite{Abdurashitov:2009tn};
the signal of short-baseline
$\bar\nu_{\mu}\to\bar\nu_{e}$
oscillations observed in the LSND experiment
\cite{Athanassopoulos:1995iw,Aguilar:2001ty}.

These results can be explained by extending the standard framework of
three-neutrino mixing with the introduction of new massive neutrinos
that in the flavor basis correspond to
sterile neutrinos
\cite{Pontecorvo:1968fh},
which do not have standard weak interactions.
The simplest scheme is that called ``3+1'',
in which there are the three standard active neutrinos
$\nu_{e}$,
$\nu_{\mu}$,
$\nu_{\tau}$,
and one new sterile neutrino
$\nu_{s}$.
In the mass basis,
there are the three standard massive neutrinos
$\nu_{1}$,
$\nu_{2}$,
$\nu_{3}$,
whose two independent squared-mass differences
correspond to
$\Delta{m}^2_{\text{SOL}}$
and
$\Delta{m}^2_{\text{ATM}}$,
and an additional massive neutrino $\nu_{4}$
which generates the short-baseline squared-mass difference
$\Delta{m}^2_{\text{41}} \simeq \Delta{m}^2_{\text{42}} \simeq \Delta{m}^2_{\text{43}} \simeq \Delta{m}^2_{\text{SBL}}$
(with the usual definition
$\Delta{m}^2_{jk} = m_{j}^2 - m_{k}^2$,
where
$m_{k}$ is the mass of $\nu_{k}$).

However,
it is possible that there are more than one
new massive neutrinos which generate more than one
$\Delta{m}^2_{\text{SBL}} \gtrsim 1 \, \text{eV}^2$.
In the literature one can find studies of
3+2
\cite{Sorel:2003hf,Karagiorgi:2006jf,Maltoni:2007zf,Karagiorgi:2009nb,Donini:2012tt},
3+3
\cite{Maltoni:2007zf},
3+1+1
\cite{Nelson:2010hz,Fan:2012ca,Kuflik:2012sw,Huang:2013zga},
and
1+3+1
\cite{Kopp:2011qd,Kopp:2013vaa}
schemes.
It turns out that in all these schemes there is
a tension between the results of
short-baseline appearance and disappearance experiments
\cite{Okada:1996kw,Bilenky:1996rw,Bilenky:1998dt,Bilenky:1999ny,Giunti:2000ur,Peres:2000ic,Grimus:2001mn,Maltoni:2002xd,Sorel:2003hf,Maltoni:2004ei,Maltoni:2007zf,GonzalezGarcia:2007ib,Karagiorgi:2009nb,Akhmedov:2010vy,Kopp:2011qd,Giunti:2011gz,Giunti:2011hn,Giunti:2011cp,Donini:2012tt,Conrad:2012qt,Archidiacono:2012ri,Archidiacono:2013xxa,Kopp:2013vaa,Giunti:2013aea}.
The reason of this tension is well understood
in the framework of the 3+1 mixing scheme
from the analytic relation which
connects the amplitudes of appearance and disappearance short-baseline
oscillations
\cite{Okada:1996kw,Bilenky:1996rw}.
In this paper we extend this explanation
to the general case of
3+$N_{s}$ mixing,
in which there are $N_{s}$ mostly sterile massive neutrinos at the eV scale
(some of which can also be at a larger scale).

For the study of neutrino oscillations in vacuum it is convenient to use
the following general expression of the probability of
$\nua{\alpha}\to\nua{\beta}$
oscillations
\cite{Bilenky:2012zp,Bilenky:2015xwa}:
\begin{align}
P_{\nua{\alpha}\to\nua{\beta}}
=
\null & \null
\delta_{\alpha\beta}
-
4
\sum_{k \neq p}
|U_{\alpha k}|^2
\left( \delta_{\alpha\beta} -  |U_{\beta k}|^2 \right)
\sin^{2}\Delta_{kp}
\nonumber
\\
\null & \null
+
8
\sum_{\stackrel{\scriptstyle j>k}{\scriptstyle j,k \neq p}}
\left|
U_{\alpha j}
U_{\beta j}
U_{\alpha k}
U_{\beta k}
\right|
\sin\Delta_{kp}
\sin\Delta_{jp}
\nonumber
\\
\null & \null
\hspace{2cm}
\times
\cos(
\Delta_{jk}
\stackrel{(+)}{-}
\eta_{\alpha\beta jk}
)
,
\label{221}
\end{align}
where
\begin{equation}
\Delta_{kp} = \frac{ \Delta{m}^{2}_{kp} L }{ 4 E }
,
\qquad
\eta_{\alpha\beta jk}
=
\operatorname{arg}\!\left[
U_{\alpha j}^{*}
U_{\beta j}
U_{\alpha k}
U_{\beta k}^{*}
\right]
,
\label{222}
\end{equation}
and $p$ is an arbitrary fixed index,
which can be chosen in the most convenient way
depending on the case under consideration.
In the case of three-neutrino mixing,
there is only one interference term in Eq.~(\ref{221}),
because for any choice of $p$ there is only one possibility for $j$ and $k$
such that $j>k$.

In the following we take into account that
the non-standard massive neutrinos must be mostly sterile,
i.e.
\begin{equation}
|U_{\alpha k}|^2 \ll 1
\qquad
(\alpha=e,\mu,\tau; \quad k=4, \ldots, N)
,
\label{smallmix}
\end{equation}
in order not to spoil the successful three-neutrino mixing explanation of
solar, atmospheric and long-baseline
neutrino oscillation measurements.
The non-standard massive neutrinos are conveniently labeled in order of increasing mass:
$m_{4} \leq m_{5} \leq \ldots \leq m_{N}$.

\begin{figure}
\centering
\includegraphics*[viewport=59 25 652 574,width=\linewidth]{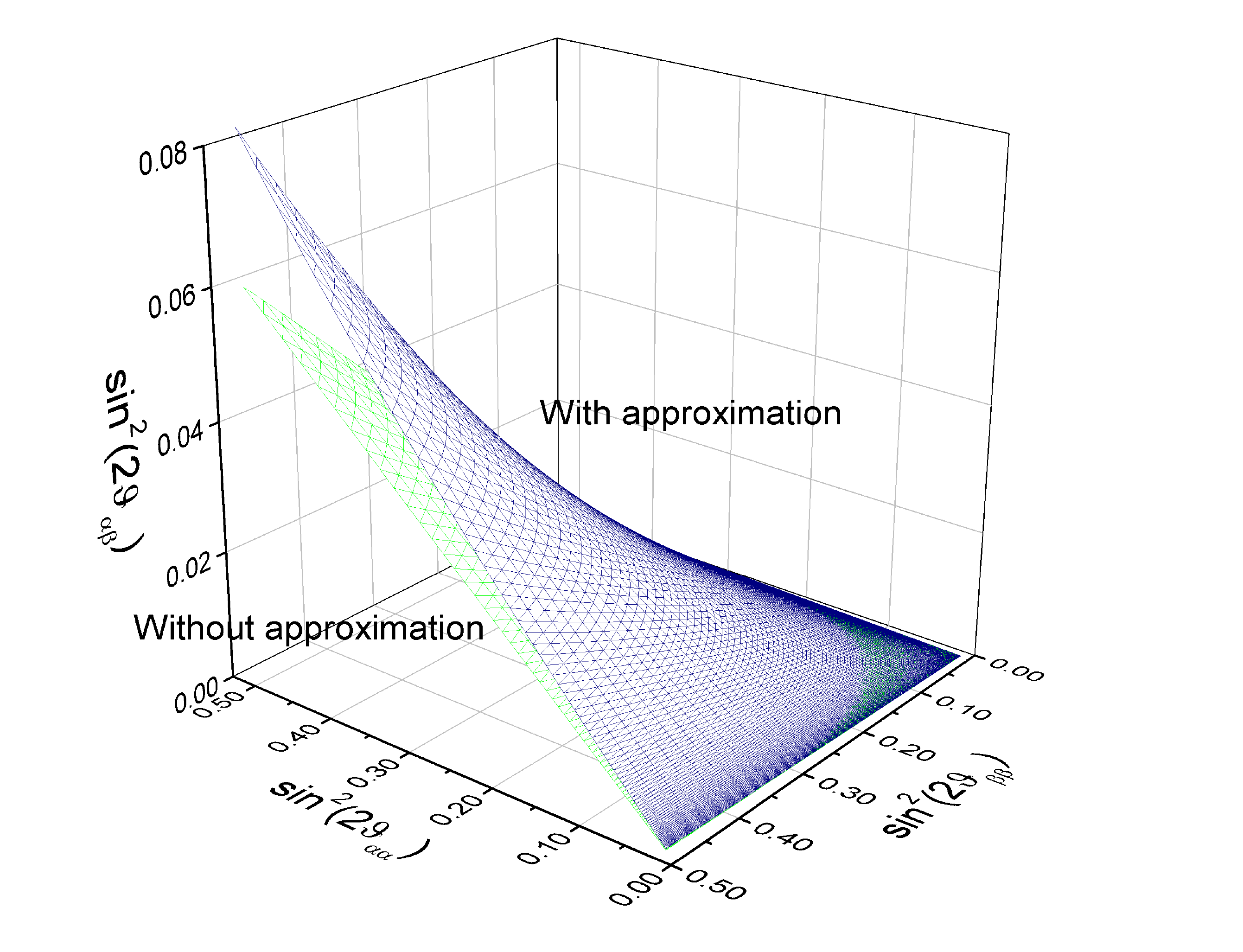}
\caption{
\label{3d}
Three-dimensional illustration of the relation in Eq.~(\ref{adex})
(without approximation)
and the approximated relation in Eq.~(\ref{adap})
(with approximation).
}
\end{figure}

\begin{figure}
\centering
\includegraphics*[width=\linewidth]{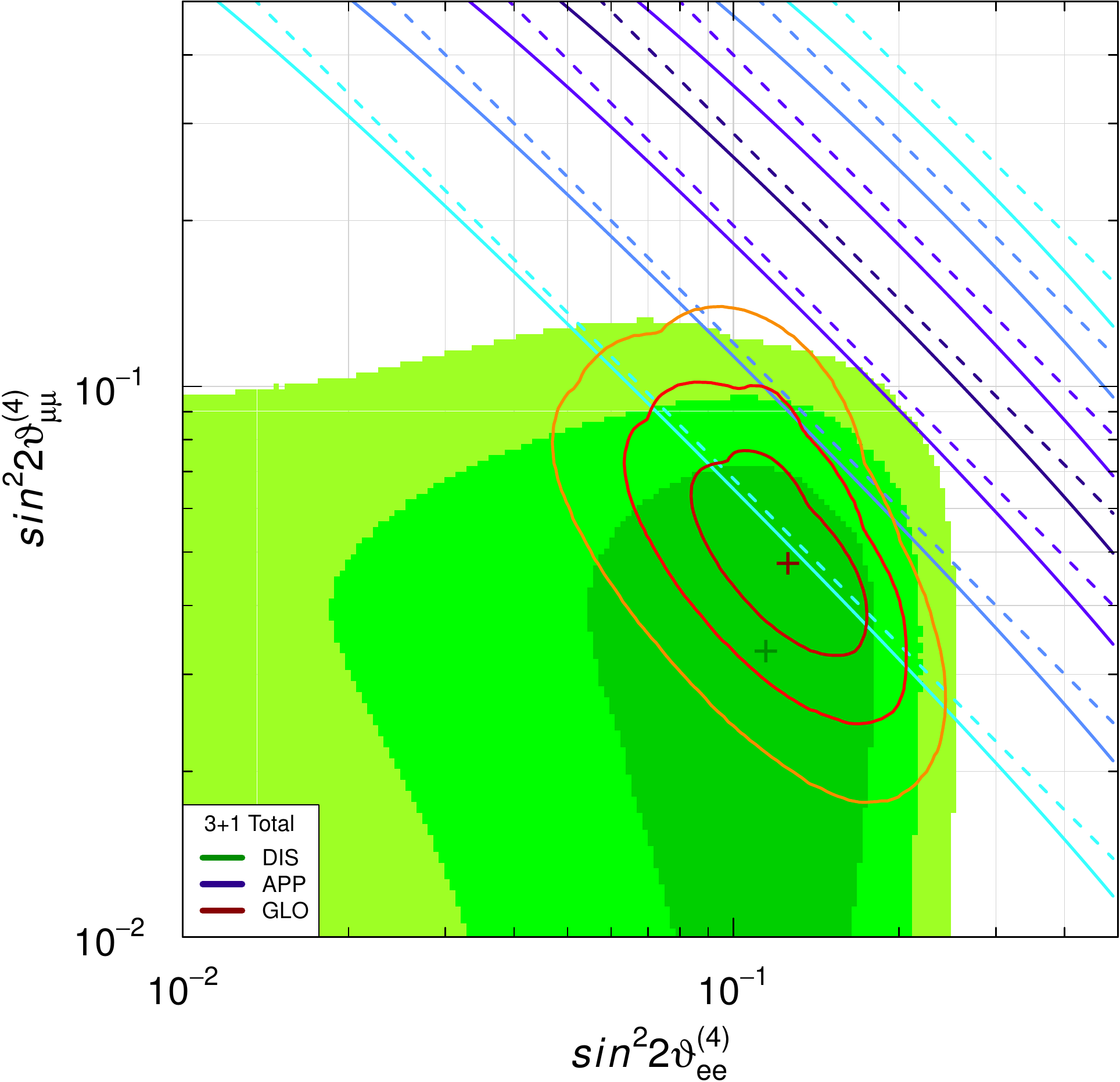}
\caption{
\label{appdis-3p1-low}
Allowed regions in the
$\sin^2 2\vartheta_{ee}^{(4)}$--$\sin^2 2\vartheta_{\mu\mu}^{(4)}$
plane
obtained in the total 3+1 fit of short-baseline data
corresponding to the 3+1-LOW fit of Ref.~\cite{Giunti:2013aea}.
The green shadowed regions are the regions allowed at
$1\sigma$,
$2\sigma$ and
$3\sigma$
by the analysis of short-baseline disappearance (DIS) data,
with the best fit value indicated by a dark-green cross.
The strips enclosed by the blue diagonal lines are allowed at
$1\sigma$,
$2\sigma$ and
$3\sigma$
by the analysis of short-baseline appearance (APP) data,
with the central best fit dark-blue line.
The solid lines correspond to the exact relation in Eq.~(\ref{adex}),
whereas the dashed lines correspond to the approximated relation in Eq.~(\ref{adap}).
The regions inside the red-orange closed curves are allowed at
$1\sigma$,
$2\sigma$ and
$3\sigma$
by the global (GLO) analysis of short-baseline data,
with the best fit value indicated by a dark-red cross.
}
\end{figure}

\begin{figure}
\centering
\includegraphics*[width=\linewidth]{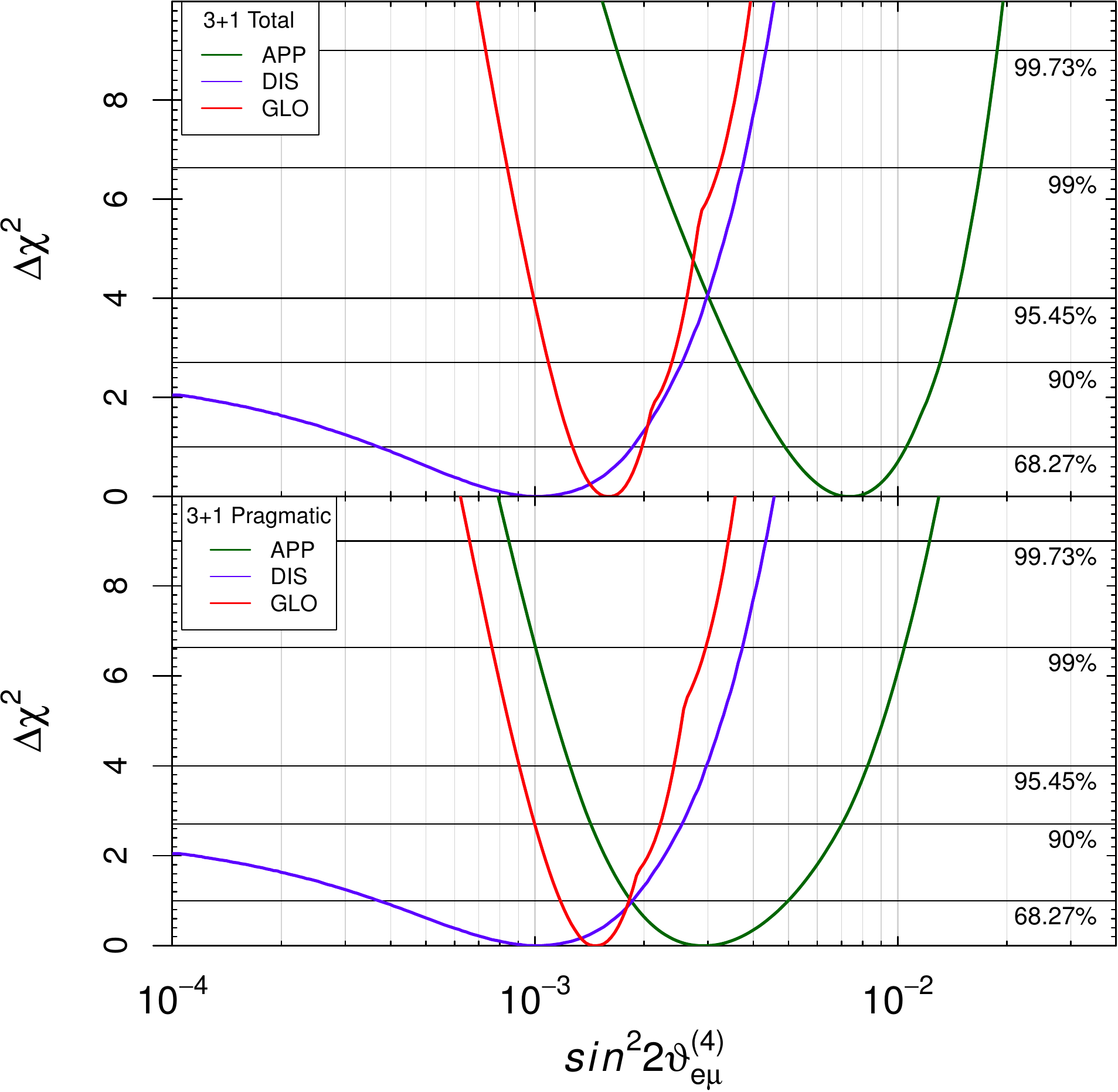}
\caption{
\label{semchi-3p1}
Marginal $\Delta\chi^2$ as a function of $\sin^2 2\vartheta_{e\mu}^{(4)}$
obtained in the total (upper panel) and pragmatic (lower panel)
3+1 analyses of short-baseline data
presented in Ref.~\cite{Giunti:2013aea}.
The horizontal lines gives the values of $\Delta\chi^2$
corresponding to
68.27\% C.L. ($1\sigma$),
90\% C.L.,
95.45\% C.L. ($2\sigma$),
99\% C.L. and
99.73\% C.L. ($3\sigma$)
for one degree of freedom.
The curves obtained from disappearance (DIS) data
are equal in the two panels.
}
\end{figure}

\begin{figure}
\centering
\includegraphics*[width=\linewidth]{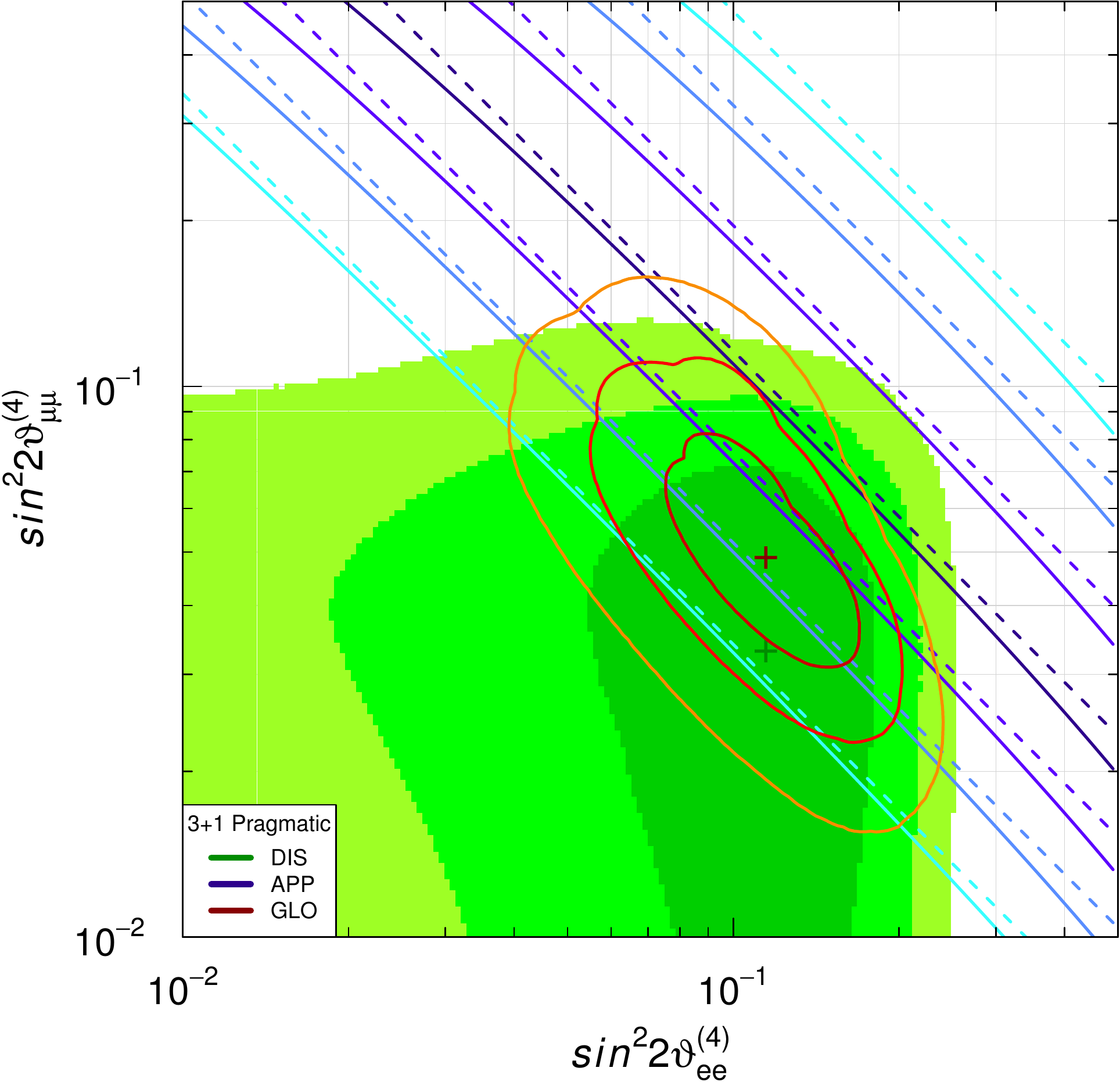}
\caption{
\label{appdis-3p1-hig}
Allowed regions in the
$\sin^2 2\vartheta_{ee}^{(4)}$--$\sin^2 2\vartheta_{\mu\mu}^{(4)}$
plane
obtained in the pragmatic 3+1 fit of short-baseline data
corresponding to the 3+1-HIG fit of Ref.~\cite{Giunti:2013aea}.
See the caption of Fig.~\ref{appdis-3p1-low}.
The green shadowed regions obtained from disappearance (DIS) data
are the same of Fig.~\ref{appdis-3p1-low}.
}
\end{figure}

\begin{figure}
\centering
\includegraphics*[width=\linewidth]{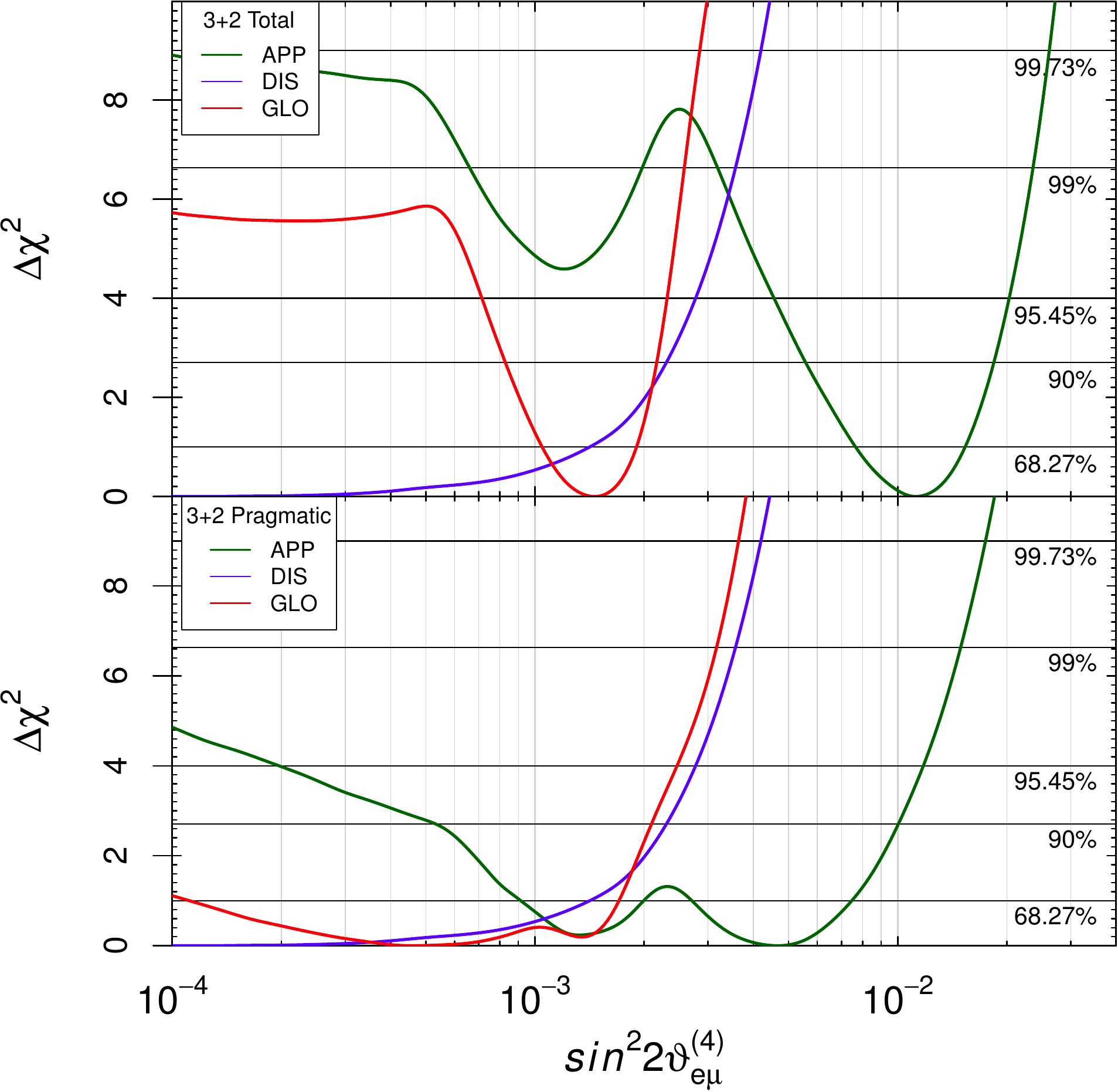}
\caption{
\label{semchi4-3p2}
Marginal $\Delta\chi^2$ as a function of $\sin^2 2\vartheta_{e\mu}^{(4)}$
obtained in the total (upper panel) and pragmatic (lower panel) 3+2 fits of short-baseline data
presented in Ref.~\cite{Giunti:2013aea}.
See the caption of Fig.~\ref{semchi-3p1}.
}
\end{figure}

\begin{figure}
\centering
\includegraphics*[width=\linewidth]{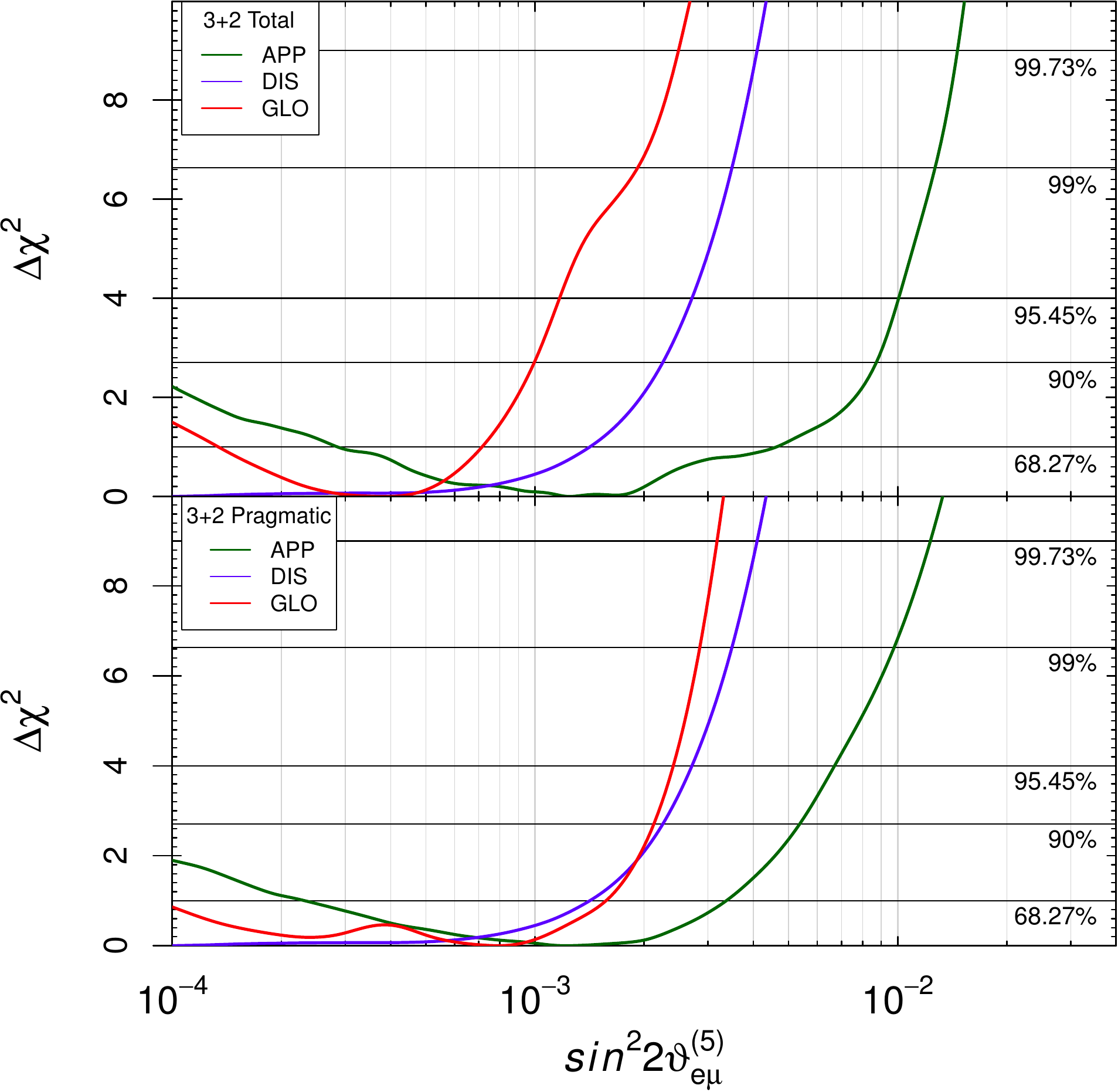}
\caption{
\label{semchi5-3p2}
Marginal $\Delta\chi^2$ as a function of $\sin^2 2\vartheta_{e\mu}^{(5)}$
obtained in the total (upper panel) and pragmatic (lower panel) 3+2 fits of short-baseline data
presented in Ref.~\cite{Giunti:2013aea}.
See the caption of Fig.~\ref{semchi-3p1}.
}
\end{figure}

\begin{figure}
\centering
\includegraphics*[width=\linewidth]{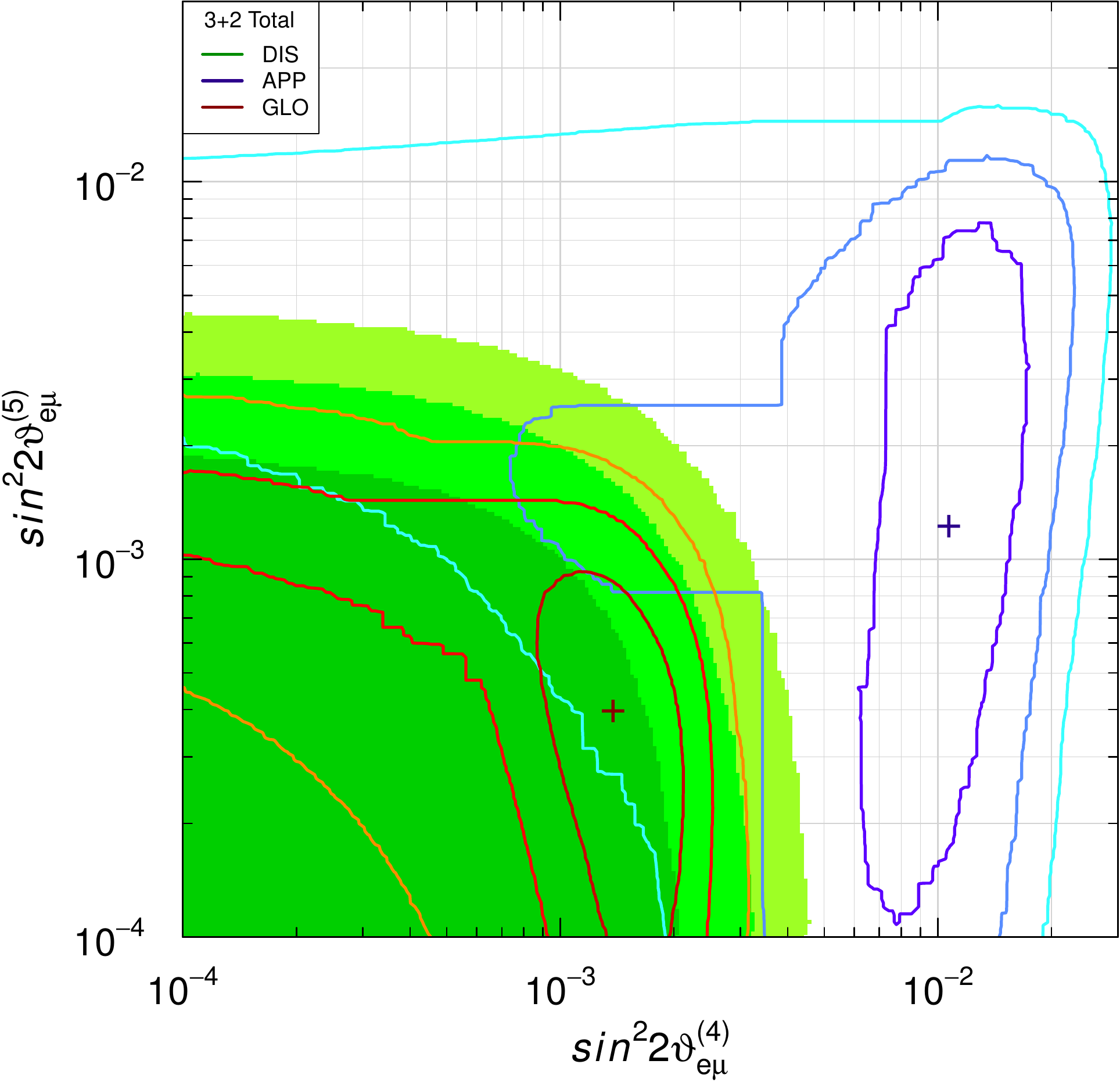}
\caption{
\label{appdis45-3p2-low}
Allowed regions in the
$\sin^2 2\vartheta_{e\mu}^{(4)}$--$\sin^2 2\vartheta_{e\mu}^{(5)}$
plane
obtained in the total 3+2 fit of short-baseline data
corresponding to the 3+2-LOW fit of Ref.~\cite{Giunti:2013aea}.
The green shadowed regions are the regions allowed at
$1\sigma$,
$2\sigma$ and
$3\sigma$
by the analysis of short-baseline disappearance (DIS) data,
with the best fit value at
$\sin^2 2\vartheta_{e\mu}^{(4)}=\sin^2 2\vartheta_{e\mu}^{(5)}=0$.
The regions inside the blue closed curves are allowed at
$1\sigma$,
$2\sigma$ and
$3\sigma$
by the analysis of short-baseline appearance (APP) data,
with the best fit value indicated by a dark-blue cross.
The regions inside the red-orange closed curves are allowed at
$1\sigma$,
$2\sigma$ and
$3\sigma$
by the global (GLO) analysis of short-baseline data,
with the best fit value indicated by a dark-red cross.
}
\end{figure}

\begin{figure}
\centering
\includegraphics*[width=\linewidth]{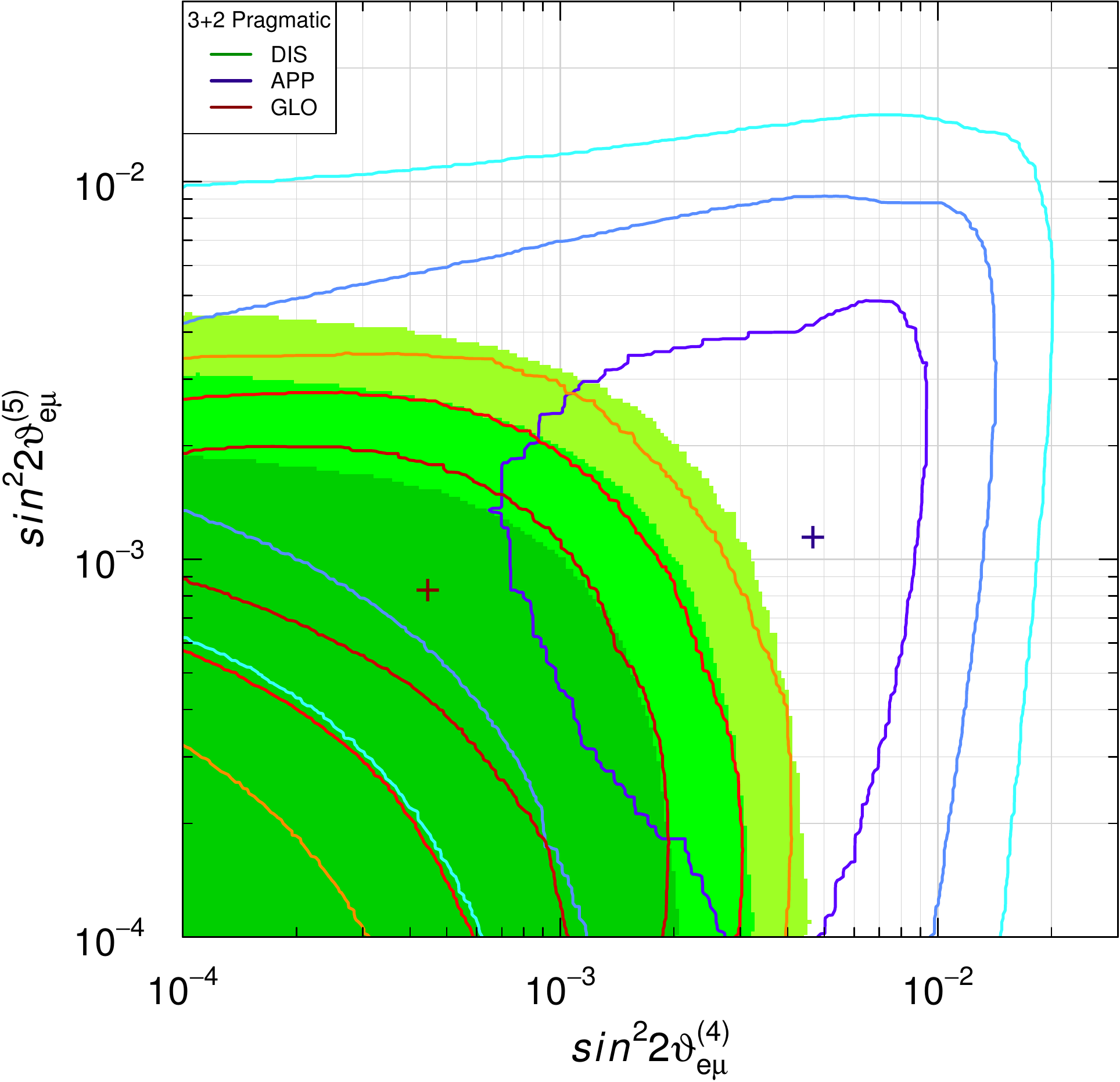}
\caption{
\label{appdis45-3p2-hig}
Allowed regions in the
$\sin^2 2\vartheta_{e\mu}^{(4)}$--$\sin^2 2\vartheta_{e\mu}^{(5)}$
plane
obtained in the pragmatic 3+2 fit of short-baseline data
corresponding to the 3+2-HIG fit of Ref.~\cite{Giunti:2013aea}.
See the caption of Fig.~\ref{appdis45-3p2-low}.
The green shadowed regions obtained from disappearance (DIS) data
are the same of Fig.~\ref{appdis45-3p2-low}.
}
\end{figure}

We are interested in the effective oscillation probabilities
in short-baseline experiments,
for which
$\Delta_{21} \ll \Delta_{31} \ll 1$.
In the general case of
3+$N_{s}$
mixing with
$\Delta{m}^2_{k1} \gtrsim \Delta{m}^2_{\text{SBL}}$
and
$\Delta_{k1} \gtrsim 1$
for $k\geq4$,
choosing $p=1$ in Eq.~(\ref{222}) we obtain
\begin{align}
P_{\nua{\alpha}\to\nua{\beta}}^{(\text{SBL})}
\simeq
\null & \null
\delta_{\alpha\beta}
-
4
\sum_{k=4}^{N}
|U_{\alpha k}|^2
\left( \delta_{\alpha\beta} -  |U_{\beta k}|^2 \right)
\sin^{2}\Delta_{k1}
\nonumber
\\
\null & \null
+
8
\sum_{k=4}^{N}
\sum_{j=k+1}^{N}
\left|
U_{\alpha j}
U_{\beta j}
U_{\alpha k}
U_{\beta k}
\right|
\sin\Delta_{k1}
\sin\Delta_{j1}
\nonumber
\\
\null & \null
\hspace{2.5cm}
\times
\cos(
\Delta_{jk}
\stackrel{(+)}{-}
\eta_{\alpha\beta jk}
)
.
\label{223}
\end{align}

Considering the survival probabilities of active neutrinos,
let us define the effective amplitudes
\begin{equation}
\sin^2 2\vartheta_{\alpha\alpha}^{(k)}
=
4 |U_{\alpha k}|^2 \left( 1 - |U_{\alpha k}|^2 \right)
\simeq
4 |U_{\alpha k}|^2
,
\label{asurv}
\end{equation}
for
$\alpha=e,\mu,\tau$
and
$k\geq4$.
The approximation is due to the constraint (\ref{smallmix}),
which allows to neglect the quadratically suppressed contribution proportional to $|U_{\alpha k}|^4$.
Dropping the quadratically suppressed terms also in the survival probabilities,
we obtain
\begin{equation}
P_{\nua{\alpha}\to\nua{\alpha}}^{(\text{SBL})}
\simeq
1
-
\sum_{k=4}^{N}
\sin^2 2\vartheta_{\alpha\alpha}^{(k)}
\sin^{2}\Delta_{k1}
,
\label{psurv}
\end{equation}
for $\alpha=e,\mu,\tau$.
Hence,
each effective mixing angle
$\vartheta_{\alpha\alpha}^{(k)}$
parameterizes the disappearance of $\nua{\alpha}$
due to its mixing with $\nua{k}$ for $k \geq 4$.

Let us now consider the probabilities of short-baseline
$\nua{\alpha}\to\nua{\beta}$
transitions between two different active neutrinos
or an active and a sterile neutrino.
We define the transition amplitudes
\begin{equation}
\sin^2 2\vartheta_{\alpha\beta}^{(k)}
=
4 |U_{\alpha k}|^2 |U_{\beta k}|^2
,
\label{atran}
\end{equation}
for
$\alpha\neq\beta$ and $k\geq4$,
which allow us to write the transition probabilities as
\begin{align}
P_{\nua{\alpha}\to\nua{\beta}}^{(\text{SBL})}
\simeq
\null & \null
\sum_{k=4}^{N}
\sin^2 2\vartheta_{\alpha\beta}^{(k)}
\sin^{2}\Delta_{k1}
\nonumber
\\
\null & \null
+
2
\sum_{k=4}^{N}
\sum_{j=k+1}^{N}
\sin 2\vartheta_{\alpha\beta}^{(k)}
\sin 2\vartheta_{\alpha\beta}^{(j)}
\sin\Delta_{k1}
\sin\Delta_{j1}
\nonumber
\\
\null & \null
\hspace{2.5cm}
\times
\cos(
\Delta_{jk}
\stackrel{(+)}{-}
\eta_{\alpha\beta jk}
)
.
\label{ptran}
\end{align}
From the first line
one can see that
each effective mixing angle
$\vartheta_{\alpha\beta}^{(k)}$
parameterizes the amount of
$\nua{\alpha}\to\nua{\beta}$
transitions
due to the mixing of $\nua{\alpha}$ and $\nua{\beta}$ with $\nua{k}$ for $k \geq 4$.
The second line in Eq.~(\ref{ptran})
is the interference between the $\nua{k}$ and $\nua{j}$
contributions for $k,j \geq 4$,
which depends on the same effective mixing angles.

Considering now the transitions between two different active neutrinos,
from Eqs.~(\ref{asurv}) and (\ref{atran})
one can see that for each value of $k\geq4$
the transition amplitude
$\sin 2\vartheta_{\alpha\beta}^{(k)}$
and
the disappearance amplitudes
$\sin 2\vartheta_{\alpha\alpha}^{(k)}$
and
$\sin 2\vartheta_{\beta\beta}^{(k)}$
depend only on the elements in $k^{\text{th}}$ column of the mixing matrix
and are related by
\begin{align}
\sin^2 2\vartheta_{\alpha\beta}^{(k)}
=
\null & \null
\left(
1
-
\sqrt{1 - \sin^2 2\vartheta_{\alpha\alpha}^{(k)}}
\right)
\nonumber
\\
\null & \null
\times
\left(
1
-
\sqrt{1 - \sin^2 2\vartheta_{\beta\beta}^{(k)}}
\right)
,
\label{adex}
\end{align}
for
$\alpha=e,\mu,\tau$.
Taking into account the constraint (\ref{smallmix}),
we have
\begin{equation}
\sin^2 2\vartheta_{\alpha\beta}^{(k)}
\simeq
\frac{1}{4}
\,
\sin^2 2\vartheta_{\alpha\alpha}^{(k)}
\,
\sin^2 2\vartheta_{\beta\beta}^{(k)}
,
\label{adap}
\end{equation}
as illustrated in
Fig.~\ref{3d}.
This relation was derived in the case of 3+1 mixing and $k=4$
in Refs.~\cite{Okada:1996kw,Bilenky:1996rw}.
Now we see that the same relation is valid in the general case of
3+$N_{s}$ mixing
between the amplitudes of appearance and disappearance oscillations
due to each squared-mass difference
$\Delta{m}^2_{k1}$
generated by a non-standard mostly sterile massive neutrino at the eV scale
(or at a larger scale).
Note that the relation is very powerful in the global fit of
short-baseline oscillation data,
because it constrain the appearance and disappearance amplitudes
for any fixed value of the corresponding
squared-mass difference.

The relation (\ref{adex}) is very important,
because it constraints the oscillation signals that can be observed in
short-baseline appearance and disappearance experiments
in any 3+$N_{s}$ mixing scheme with sterile neutrinos.
Its experimental test is crucial for the acceptance or rejection of
these schemes.
In particular,
since both
$\sin^2 2\vartheta_{ee}^{(k)}$
and
$\sin^2 2\vartheta_{\mu\mu}^{(k)}$
are small,
the amplitudes of short-baseline
$\nua{\mu}\leftrightarrows\nua{e}$
transitions
implied by the results of disappearance experiments are quadratically suppressed
and in tension with the short-baseline
$\bar\nu_{\mu}\to\bar\nu_{e}$
transitions
observed in the LSND experiment
\cite{Okada:1996kw,Bilenky:1996rw,Bilenky:1998dt,Bilenky:1999ny,Giunti:2000ur,Peres:2000ic,Grimus:2001mn,Maltoni:2002xd,Sorel:2003hf,Maltoni:2004ei,Maltoni:2007zf,GonzalezGarcia:2007ib,Karagiorgi:2009nb,Akhmedov:2010vy,Kopp:2011qd,Giunti:2011gz,Giunti:2011hn,Giunti:2011cp,Donini:2012tt,Conrad:2012qt,Archidiacono:2012ri,Archidiacono:2013xxa,Kopp:2013vaa,Giunti:2013aea}.

In the following we illustrate the importance of the constraint (\ref{adex})
(and its approximation (\ref{adap}))
in the cases of
3+1
and
3+2
mixing
using the results of the fits of short-baseline neutrino oscillation data
presented in Ref.~\cite{Giunti:2013aea}.
We denote with the labels ``APP'' and ``DIS''
the fits of appearance and disappearance data alone,
respectively.
The label ``GLO''
denotes the global combined fit of appearance and disappearance data.
Furthermore,
we distinguish between ``total''
and ``pragmatic'' fits.
A total fit
corresponds to a ``LOW'' fit in Ref.~\cite{Giunti:2013aea},
in which all the data of short-baseline neutrino oscillation experiments
considered in Ref.~\cite{Giunti:2013aea} are taken into account.
A pragmatic fit
corresponds to a ``HIG'' fit in Ref.~\cite{Giunti:2013aea},
in which all the data of short-baseline neutrino oscillation experiments
considered in Ref.~\cite{Giunti:2013aea} are taken into account
except the anomalous low-energy bins of the MiniBooNE experiment
\cite{AguilarArevalo:2008rc,Aguilar-Arevalo:2013pmq}.
This ``pragmatic approach'' was advocated
in Ref.~\cite{Giunti:2013aea}
because the anomalous low-energy bins of the MiniBooNE experiment
\cite{AguilarArevalo:2008rc,Aguilar-Arevalo:2013pmq}
are incompatible with neutrino oscillations
and may be due to background
(this problem is going to be investigated in the MicroBooNE experiment
\cite{MicroBooNE-2007,Szelc:2015dga}).
Note that only the APP and GLO fits are different
in the total and a pragmatic fits,
whereas the DIS fit of disappearance data is the same.

Figure~\ref{appdis-3p1-low}
shows the regions
in the
$\sin^2 2\vartheta_{ee}^{(4)}$--$\sin^2 2\vartheta_{\mu\mu}^{(4)}$
plane
allowed at
$1\sigma$,
$2\sigma$ and
$3\sigma$
by the total 3+1 fit of short-baseline data
presented in Ref.~\cite{Giunti:2013aea}
(corresponding to the ``3+1-LOW'' fit in Ref.~\cite{Giunti:2013aea}).
One can see that there is a clear tension between the
regions allowed by disappearance and appearance data.
The relation (\ref{adex}) allows us to illustrate the tension
also by showing the
marginal $\Delta\chi^2$ as a function of $\sin^2 2\vartheta_{e\mu}^{(4)}$
in the upper panel of Fig.~\ref{semchi-3p1}.
One can see that there is no overlap of the
intervals of $\sin^2 2\vartheta_{e\mu}^{(4)}$
allowed by the appearance and disappearance data
at less than about $2\sigma$
and the result of the global fit is a compromise
with the best-fit dominated by the disappearance data.

The fit of the anomalous MiniBooNE low-energy bins with 3+1 neutrino oscillations
requires a small value of $\Delta{m}^2_{41}$
and a large value of $\sin^22\vartheta_{e\mu}^{(4)}$,
which are in strong tension with the data of disappearance experiments
\cite{Giunti:2011hn,Giunti:2011cp,Giunti:2013aea}.
Hence,
we can alleviate the appearance-disappearance tension by adopting the
pragmatic approach
described above.
The allowed regions in the
$\sin^2 2\vartheta_{ee}^{(4)}$--$\sin^2 2\vartheta_{\mu\mu}^{(4)}$
plane
in Fig.~\ref{appdis-3p1-hig}
obtained with the pragmatic 3+1 fit of short-baseline data
show that the appearance-disappearance tension is acceptable.
A comparison with Fig.~\ref{appdis-3p1-low}
shows that the strips allowed by the analysis of short-baseline appearance (APP) data
have moved towards smaller values of
$\sin^2 2\vartheta_{ee}^{(4)}$ and $\sin^2 2\vartheta_{\mu\mu}^{(4)}$,
which are in agreement with the analysis of disappearance data.
The same behavior can be observed by confronting the two panels
in Fig.~\ref{semchi-3p1},
where one can see that in the pragmatic approach
the intervals of $\sin^2 2\vartheta_{e\mu}^{(4)}$
allowed by the appearance and disappearance data
overlap at about $1\sigma$.

Let us now consider the 3+2 mixing scheme.
Figures~\ref{semchi4-3p2} and \ref{semchi5-3p2}
show the marginal $\Delta\chi^2$ as a function of
$\sin^2 2\vartheta_{e\mu}^{(4)}$ and $\sin^2 2\vartheta_{e\mu}^{(5)}$,
respectively.
They are different because the ordering
$\Delta{m}^2_{41} \leq \Delta{m}^2_{51}$
breaks the degeneracy between the mixing of $\nu_{4}$ and $\nu_{5}$.
Thanks to the relation (\ref{adex}),
these figures are sufficient to visualize the appearance-disappearance tension.

The upper panel in Fig.~\ref{semchi4-3p2}
shows that in the total 3+2 fit there is no overlap of the
intervals of $\sin^2 2\vartheta_{e\mu}^{(4)}$
allowed by the appearance and disappearance data
at more than $2\sigma$,
which corresponds to an appearance-disappearance tension which is worse
than that in the total 3+1 fit illustrated in the upper panel of
Fig.~\ref{semchi-3p1}
(see the discussion in Ref.\cite{Archidiacono:2013xxa}).
On the other hand,
the upper panel in Fig.~\ref{semchi5-3p2} shows that
the data do not constrain much the value of
$\sin^2 2\vartheta_{e\mu}^{(5)}$
and the corresponding marginal $\Delta\chi^2$'s do not show
an appearance-disappearance tension.
However,
it is clear that the manifestation of an appearance-disappearance tension
in one of the oscillation amplitudes is enough.

The secondary local minimum at
$\sin^2 2\vartheta_{e\mu}^{(4)} \approx 1.2 \times 10^{-3}$
of the appearance marginal $\Delta\chi^2$
in the upper panel in Fig.~\ref{semchi4-3p2}
is due to the possibility to fit the appearance data with
$
\sin^2 2\vartheta_{e\mu}^{(4)}
\approx
\sin^2 2\vartheta_{e\mu}^{(5)}
\approx
1-2 \times 10^{-3}
$
(with
$\Delta{m}^2_{41} \approx 2 \, \text{eV}^2$
and
$\Delta{m}^2_{51} \approx 4 \, \text{eV}^2$).
This can be seen from the correlated allowed regions in the
$\sin^2 2\vartheta_{e\mu}^{(4)}$--$\sin^2 2\vartheta_{e\mu}^{(5)}$
plane shown in Fig.~\ref{appdis45-3p2-low}
in which the $2\sigma$ region allowed by appearance data has a bulge
for
$
\sin^2 2\vartheta_{e\mu}^{(4)}
\approx
\sin^2 2\vartheta_{e\mu}^{(5)}
\approx
1-2 \times 10^{-3}
$
(this type of figure has been already presented in Ref.~\cite{Giunti:2011gz};
similar figures for definite values of $\Delta{m}^2_{41}$ and $\Delta{m}^2_{51}$
have been presented in Ref.~\cite{Kopp:2013vaa}).
From this figure one can have another view of
the appearance-disappearance tension,
which is clearly dominated by $\sin^2 2\vartheta_{e\mu}^{(4)}$.

The lower panels of Figs.~\ref{semchi4-3p2} and \ref{semchi5-3p2}
and Fig.~\ref{appdis45-3p2-hig}
illustrate the appearance-disappearance tension in the pragmatic 3+2 fit.
One can clearly see that, as in the 3+1 case, it is much milder than that
obtained in the total fit,
indicating an acceptable fit of the data.
The secondary local minimum at
$\sin^2 2\vartheta_{e\mu}^{(4)} \approx 1.2 \times 10^{-3}$
of the appearance marginal $\Delta\chi^2$
in the lower panel in Fig.~\ref{semchi4-3p2}
is due to the same reasons as that in the upper panel explained above
and corresponds to the swelling
towards
$
\sin^2 2\vartheta_{e\mu}^{(4)}
\approx
\sin^2 2\vartheta_{e\mu}^{(5)}
\approx
1-2 \times 10^{-3}
$
of the $1\sigma$ region allowed by appearance data
in Fig.~\ref{appdis45-3p2-hig}.

In conclusion,
in this paper we have derived the relation (\ref{adex})
(and its approximation (\ref{adap}))
between the
amplitudes of short-baseline appearance and disappearance oscillations
in 3+$N_{s}$ neutrino mixing schemes.
This relation is the origin of the appearance-disappearance tension
that is found from the analysis of the existing data
in any 3+$N_{s}$ neutrino mixing scheme.
We have illustrated the power of the relation (\ref{adex})
to reveal the appearance-disappearance tension
in the cases of
3+1
and
3+2
mixing
using the results of the fits of short-baseline neutrino oscillation data
presented in Ref.~\cite{Giunti:2013aea}.

\begin{acknowledgments}
This work
was partially supported by the research grant {\sl Theoretical Astroparticle Physics} number 2012CPPYP7 under the program PRIN 2012 funded by the Ministero dell'Istruzione, Universit\`a e della Ricerca (MIUR).
E. Z. thanks the support of funding grants 2013/02518-7 and 2014/23980-3, S\~ao Paulo Research Foundation (FAPESP).
\end{acknowledgments}


\end{document}